# The Future of Food: How Artificial Intelligence is Transforming Food Manufacturing

Xu Zhou[1,2*], Ivor Prado[2*], AIFPDS participants, Ilias Tagkopoulos[1,2]

[1]Department of Computer Science & Genome Center, University of California, Davis
[2]USDA/NSF AI Institute for Next Generation Food Systems

**Correspondence:** Ilias Tagkopoulos (itagkopoulos@ucdavis.edu)
*Equal contribution

# THE FUTURE OF FOOD:
# How Artificial Intelligence is Transforming Food Manufacturing

AI for Food Product Development Sympisium • October 13 2025 •
• AI Institute for Next-Generation Food Systems (AIFS) UC, Davis

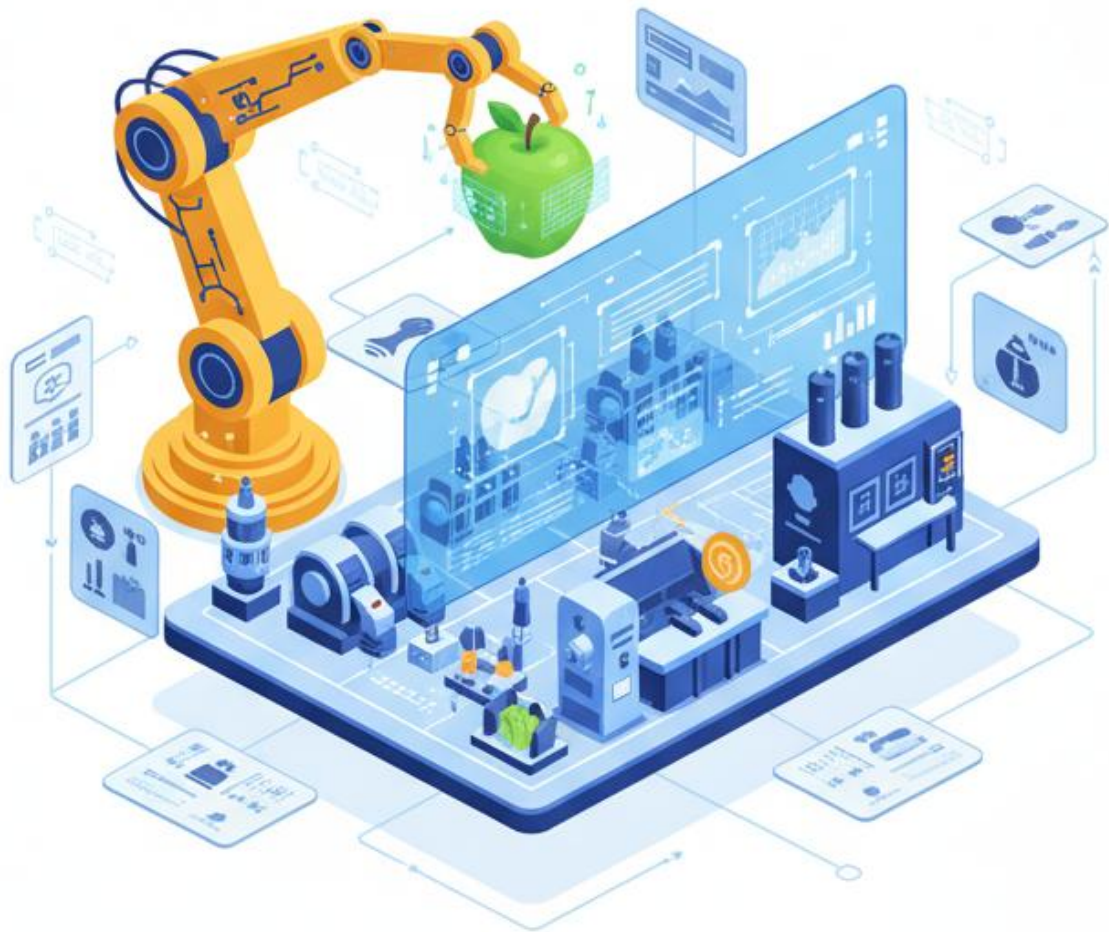

# Table of Contents

**Executive Summary**

The AI for Food Product Development Symposium (AIFPDS), hosted by AI Institute for Next Generation Food Systems (AIFS) at the University of California, Davis, brought together leaders in food science, engineering, and computer science from academia, industry, and government to explore how AI is transforming food manufacturing. The event marked an important step in reimagining how foods are formulated, processed, distributed, and perceived. Participants also emphasized the importance of educating and training the next generation of food and data scientists who will lead innovation within increasingly digital and interdisciplinary environments.

The future of AI in food manufacturing will require interoperable data standards, transparent and explainable models, ethical governance, and a skilled workforce capable of integrating digital tools with scientific and practical knowledge to drive innovation responsibly. The participants discussed five areas: (1) *Supply Chain*, where AI can enable real-time monitoring of food quality and safety, traceability, predictive logistics, and advance the circular economy; (2) *Formulation and Processing*, where hybrid data-driven and physics-based models can redefine product development and process design; (3) *Consumer Insights and Sensory*, where AI-driven multimodal models can connect chemical composition, physical structure, and consumer science to predict consumer perception and preference; (4) *Nutrition and Health*, where predictive and personalized models that integrate food omics and dietary data transform preventive health care through precision nutrition; and (5) *Education and Training*, where integrating AI literacy, ethics, critical thinking, and hands-on problem-solving throughout curricula will cultivate an interdisciplinary, critically engaged workforce.

The participants also identified challenges, including uneven data sets that hinder collaboration, limited trust and governance frameworks for ethical and secure data sharing, and a persistent skills gap between food domain expertise and data science. We proposed next steps focusing on building shared data and model infrastructures supported by interoperable standards, privacy-preserving frameworks, and open benchmarks to validate AI performance. The discussions also emphasized the importance of cross-sector collaborations to accelerate the translation from research into practice and the incentive towards workforce development initiatives that integrate AI literacy and interdisciplinary problem-solving across the food system.

# Introduction

Artificial intelligence (AI) is transforming industries faster and integrating more domains when compared to previous industrial revolutions. Over the past decade, advances in machine learning, natural language processing, and multi-omics data integration have expanded our understanding of food composition, processing and their effect on human health. The merging of digital, physical, and biological innovations, often referred to as the Fourth Industrial Revolution, is also being applied to redefine every stage of the food value chain (Schwab, 2016). These advances now enable connectivity across data domains, linking agricultural production, ingredient functionality, processing parameters, sensory experiences, and consumer health outcomes within a unified, intelligent decision framework.

At the same time, demand for healthier, more sustainable, and flavorful products continue to grow, while supply chains struggle with environmental pressures, stricter sustainability regulations, and increasing logistical complexity. These challenges, combined with the inherent variability of food ingredients and faster product turnover, expose the limitations of traditional food product research and development methods, still dependent on empirical testing and limited data, which can no longer keep pace with the required speed and diversity of global market needs (Habib, M. et al., 2025). Yet despite these converging forces, AI adoption across the food system remains uneven, constrained by heterogeneous data, limited systems and model interoperability, and a persistent skills gap between data scientists and food domain experts (Tagkopoulos et al., 2022).

Addressing these challenges requires an integrated approach capable of tackling the several challenges on the food chain, such as supply chain disruptions, nutritional disparities, the growing demand for sustainable manufacturing and multidisciplinary workforce training. AI offers a unique opportunity to respond to these issues holistically, optimizing agricultural inputs, predicting product performance, and enabling smart manufacturing with reduced waste and energy use. However, realizing this potential demands more than technical innovation. It calls for shared data standards, trustworthy model governance, and a skilled workforce able to effectively transform AI insights into impactful real-world innovations.

Recognizing this opportunity, the AI Institute for Next Generation Food Systems (AIFS), one of 29 established National AI Institutes and the only one dedicated to advancing AI applications across the entire food system, brings together experts in agriculture, biology, chemistry, food science and engineering, and data science to promote the responsible and effective use of AI throughout the food value chain. To further this mission, AIFS hosted the inaugural AI for Food Product Development Symposium at the University of California, Davis, on October 13, 2025. The event convened academic and industry leaders from food processing, computer science, and related disciplines to explore how AI can transform product design, manufacturing, nutrition, and workforce development.

Insights from the symposium are the foundation of this white paper, which is organized around five thematic areas (Fig. 1) where AI can have the greatest near-term impact: 1) Supply Chain,

2) Formulation and Processing, 3) Consumer Insights and Sensory, 4) Nutrition and Health, and 5) Education and Training. Each section summarizes the key capabilities, challenges, and priorities identified by participants, outlining a roadmap for integrating AI into food manufacturing to enhance innovation, sustainability, and human well-being.

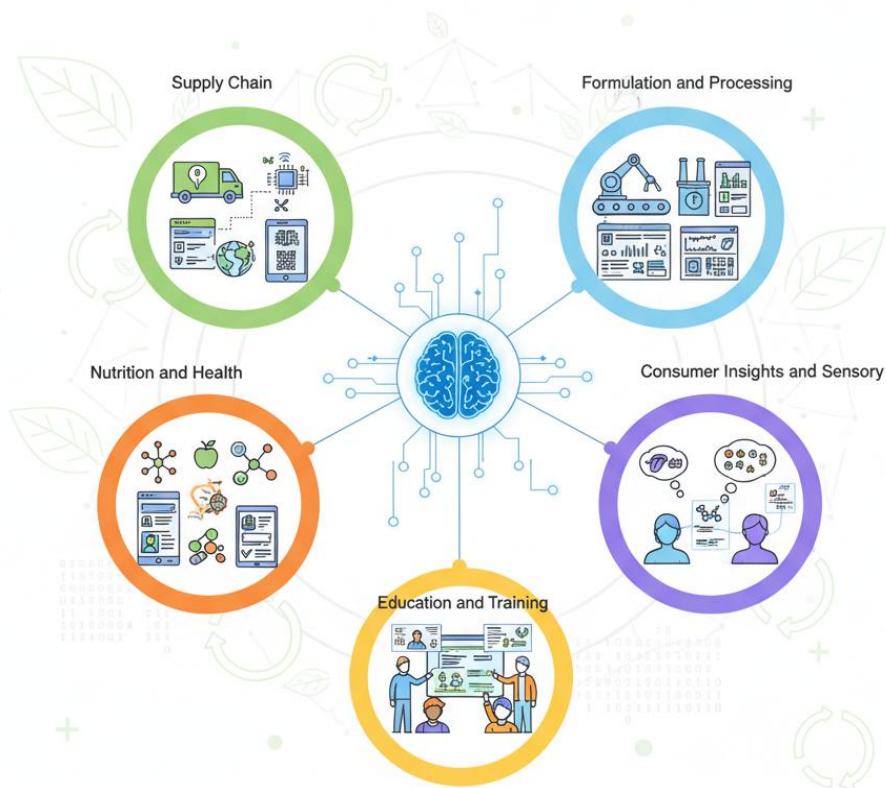

Fig. 1 AI as a central hub integrating five key domains of the future food system

## 1. Supply Chain

*1.1 Background*

A food supply chain is the system that connects farms, processors, distributors, retailers, and consumers (Zhou et al., 2025). Yet even with growing digital capacity (Rejeb et al., 2022), many supply chains remain fragmented and disconnected. Data on food origin, quality, distribution and storage conditions, and handling are collected at different points but rarely shared in real time or in compatible formats. This limits traceability, slows responses to spoilage or contamination, and reduces the overall system efficiency. Commercial competition, data privacy concerns, and uneven access to technology make cooperation difficult. As a result, food waste is estimated at between 30–40% of the food supply in the world (EPA, 2025). As global demand and climate challenges grow, building a connected and transparent supply chain has become essential for food security and sustainability.

*1.2 AI for Supply Chain Optimization*

AI-driven systems can be used to connect data across the entire food chain and support better decisions. They can be used to gather and interpret information from sensors, tracking devices, and enterprise databases to identify how food moves and changes through each stage. One recently developed AI-driven application combines physical and biological knowledge with data analytics to model how factors such as temperature, humidity, mechanical vibration, and storage time affect quality and shelf life of fresh produce. In this example, the AI models are utilized to predict when and where losses may occur, which supports decisions on adjusting transport routes, storage conditions, or inventory levels before problems arise.

To work effectively, the process begins with cleaning and aligning data from multiple sources so that all information refers to the same time, location, and product batch. AI algorithms then interpret how environmental and operational conditions influence product quality and safety. The system produces forecasts that can be examined and verified by supply chain managers, ensuring transparency and human oversight. The results guide day-to-day actions such as rerouting shipments, adjusting temperature controls, or rescheduling deliveries to reduce waste and save energy. Over time, as new data is added, the models improve, learning from real-world operations. This continuous learning loop supports a more adaptive and resilient food supply network that balances efficiency, quality, and environmental goals.

*1.3 Future Directions*

Future progress will depend on open data standards, secure collaboration, and clear scientific methods. Research should focus on combining data analytics with process-based simulations that represent heat and mass transfer and microbial growth. This approach will make predictions both accurate and understandable. Shared and standardized datasets are needed to test and validate models across commodities and regions. Privacy-preserving data-sharing methods will allow multiple organizations to contribute without exposing sensitive information. Large-scale field trials are essential to measure the true impact of AI on reducing waste, improving energy efficiency, and lowering greenhouse gas emissions. By integrating explainable, science-based computation with reliable digital infrastructure, we can build a transparent, predictive, and sustainable supply chain ready to meet future challenges in food security and climate resilience.

## 2. Formulation and Processing

*2.1 Background*

Food formulation and processing is how agricultural ingredients are selected and converted to food products with desired texture, flavor, safety, and nutrition. The current processes rely heavily on experience, trial and error, and operator intuition. However, ingredient variability, especially in newer sources such as plant proteins, creates major challenges. Two batches of the same ingredient can behave very differently, even when certificates of analysis report similar values. This inconsistency makes it difficult to predict how materials will perform during food processes, such as mixing, extrusion, and thermal processing. The lack of standardized testing methods for ingredient functionality, such as emulsification or gel strength, further limits comparability between suppliers. Data generated during processing are often incomplete, isolated, or proprietary, preventing broader learning. As a result, product development cycles remain slow, and knowledge is lost when experts retire or change roles (Bidyalakshmi et al., 2025; Khan et al., 2022; Vishnuvarthanan et al., 2026).

*2.2 AI for Formulation and Processing*

AI can help capture, organize, and interpret the large volume of data generated during food formulation and processing. It can connect information from raw material characterization, equipment settings, and product outcomes to identify how specific variables influence quality and performance. By analyzing the relationships, AI can guide formulation adjustments and help operators adapt to ingredient variability. Instead of depending solely on trial and error, manufacturers can use data-informed recommendations to achieve target texture, stability, or flavor more efficiently. AI can also transform the way process knowledge is collected and shared. Historical production data, operator notes, and sensory feedback can be digitized and linked to processing outcomes, building a knowledge base that captures decades of expertise. Machine learning can then reveal hidden relationships among parameters, such as temperature, moisture, pressure, and ingredient ratio that determine product success. These insights can shorten development time and improve reproducibility. In extrusion, for example, AI can simulate how changing raw materials or extrusion conditions will affect product structure and sensory before physical testing (PIPA, 2025). Over time, this creates a digital memory for each process, allowing new employees to learn faster and reducing dependence on individual experts. When integrated carefully, AI can support continuous improvement without replacing human creativity. It can suggest options, flag anomalies, or prompt users to consider overlooked variables rather than dictate outcomes. The most effective systems combine automation with operator judgment, creating a feedback loop that encourages experimentation while minimizing waste.

*2.3 Future Directions*

Further progress in AI-driven formulation and processing depends on three priorities: 1) better data, 2) shared standards, and 3) stronger collaboration. Standardized analytical methods for testing ingredient functionality are needed so that data from different suppliers and labs can be compared directly. Secure, interoperable data platforms will allow processors, equipment manufacturers, and ingredient companies to contribute to shared learning while protecting intellectual property. New sensing technologies and in-line monitoring tools can provide processing data that AI needs to model transformations accurately (Bowler et al., 2022). Future research should also focus on preserving and evolving process knowledge (Ding et al., 2023). Building systems that retain historical formulations, processing conditions, and outcomes will create a long-term memory for the food industry. AI models must remain interpretable and flexible, showing how input variables lead to observed results. Pilot- and industrial-scale studies that measure improvements in product consistency, innovation speed, and resource efficiency will help demonstrate impact. If implemented responsibly, AI can make food formulation and processing faster, more consistent, and more sustainable, turning individual experience into collective intelligence for the next generation of food innovation.

## 3. Consumer Insights and Sensory

*3.1 Background*

Consumer insight and sensory science study how people experience food—its flavor, texture, aroma, and appearance—and how these perceptions influence preference and behavior. These fields connect food design with consumer satisfaction. Yet collecting and interpreting sensory data is slow, subjective, and expensive. Trained sensory panels are limited in scale, and consumer testing requires large, diverse populations that are difficult to recruit and manage. Variability in language, culture, and individual perception further complicates comparisons between studies. Traditional methods rely on controlled lab settings and manual data collection, which constrain sample size and limit relevance to real-world eating experiences. Much of the data remains proprietary, preventing broader learning. As a result, sensory research often provides only partial insight into how consumers interact with food.

*3.2 AI for Consumer Insight and Sensory Research*

AI offers new ways to predict food flavor and texture as well as connect sensory perception with consumer behavior at scale (Gunning and Tagkopoulos, 2025; Nunes et al., 2023). AI can integrate data from chemistry, sensory panels, social media, and consumer feedback to map how ingredients and processing affect human perception. By analyzing patterns in language, culture, emotion, and behavior, AI can identify which sensory attributes drive liking, purchase intent, and satisfaction. These insights can guide product design to match evolving consumer expectations, regional preferences, and health priorities.

AI also accelerates sensory testing and enhances consistency across evaluations. Advanced digital tools such as the electronic nose (e-nose) and electronic tongue (e-tongue) can simulate human olfactory and gustatory perception through sensor arrays and AI-based pattern recognition (Tan and Xu, 2020). These instruments generate extensive datasets that machine learning models interpret to classify, predict, and quantify food qualities such as aroma, flavor, and freshness. AI-driven image and audio analysis can capture subtle emotional cues—such as facial expressions or voice tone—during tasting sessions, providing objective measures of affective response. Natural language processing further supports sensory research by translating and interpreting open-ended consumer feedback from multiple languages, allowing for cross-cultural comparison and global trend analysis. By linking these behavioral and perceptual data with product composition and process parameters, AI can model and predict how consumers may respond to new formulations, packaging, or sensory profiles before full-scale testing.

Importantly, AI complements rather than replaces human expertise. Sensory scientists remain essential for designing experiments, interpreting outputs, and validating models. While AI provides analytical depth, pattern discovery, and scalability, human judgment ensures scientific rigor and contextual interpretation. When used together, they enable faster, more inclusive, and more reliable insight into how consumers experience and evaluate food

*3.3 Future Directions*

Advancing AI in consumer and sensory science will require collaboration across food companies, research institutions, and data scientists. Shared and standardized sensory datasets are critical for model training and comparison. Research should focus on transparent and interpretable AI methods that clearly show how sensory attributes, ingredients, and emotions relate to consumer preferences. Integrating sensory and consumer data with nutritional and environmental factors could create a fuller understanding of what drives healthy and sustainable choices.

Future studies should test how AI can improve product development cycles, enable more personalized food experiences, and make consumer research more inclusive across cultures. Measuring how AI tools affect prediction accuracy, testing efficiency, and real-world acceptance will help demonstrate their impact. When applied responsibly, AI can turn fragmented sensory observations into integrated consumer insight, helping the food system design products that are not only appealing but also healthier, equitable, and sustainable.

## 4. Nutrition and Health

*4.1 Background*

Nutrition and health research examines how the composition of foods interacts with the human body to influence wellness and disease. This field connects food chemistry, metabolism, genetics, and public health. However, progress is slowed by several challenges. Foods that appear identical, such as two strawberries, can differ in molecular composition depending on where and how they were grown. These differences affect their nutritional and bioactive properties, yet the existing databases rarely capture this variation. Also, people eat meals composed of complex food matrices, where interactions among nutrients, additives, and preparation methods influence health effects. On the human side, people respond differently to the same foods because of genetics, gut microbiota, lifestyle, and metabolic diversity (Parizadeh and Arrieta, 2023; Vernocchi et al., 2020). This dual variability of food and of people makes it difficult to link specific foods to specific health outcomes.

Despite the existence of many food composition databases worldwide such as USDA FoodData Central, most include only macro nutrients such as proteins, fats, and carbohydrates, offering limited insight into the chemical complexity of foods. FoodAtlas, a recently developed food-chemical database, was designed to bridge this gap by extracting 230,848 food–chemical composition relationships from 155,260 scientific papers, including 106,082 (46%) that had never been reported in any previous database (Youn et al., 2024). Understanding the full spectrum of chemical constituents in foods is necessary for elucidating how dietary components influence human nutrition and health. Ongoing work aims to extend FoodAtlas to include bioactive compounds, such as phenolic acids, and flavonoids that play key roles in bioactivity and disease prevention. However, there is a challenge: the composition of foods does not necessarily reflect what reaches the human body. Many food-derived chemicals undergo digestion, metabolism, and transformation through interactions with the gut microbiota and host tissues. Current data on how these molecules are absorbed, modified, and utilized in the body or organs are limited. Devices that can track digestion and metabolism in real time are emerging, but they have yet to be widely integrated into clinical or population studies. Many studies have fragmented understanding of how diet shapes health, compounded by outdated nutritional guidance and inconsistent data standards across countries.

*4.2 AI for Nutrition and Health*

AI can help bridge these gaps by connecting molecular food data with biological and health information. AI can process large, diverse datasets from food chemistry and agricultural records to clinical, genetic, and wearable data to identify meaningful patterns across scales. By analyzing these relationships, AI can help predict how food composition affects metabolism and health in different individuals or groups. It can also simulate how small molecular changes caused by ripening, storage, or processing alter nutritional quality and bioactivity. These insights can guide agricultural and food production practices toward higher nutritional value and greater consistency.

AI can further support the design of personalized and population-level nutrition strategies. By integrating information about diet, genetics, and physiology, AI models can estimate how people with similar metabolic profiles might respond to certain foods. This makes it possible to design diets tailored to specific needs, such as managing diabetes or reducing inflammation, while maintaining affordability and cultural relevance. Visualization tools powered by AI can also help translate complex nutritional data into easy-to-understand information for consumers, empowering healthier choices. For example, improved nutrition labeling systems, informed by AI-driven analysis, could provide consistent, transparent, and globally comparable indicators of health. When applied across the food system, AI can also identify where nutritional loss occurs—from farm to processing—and suggest interventions to preserve nutrient quality. By connecting insights across production, formulation, and health outcomes, AI can shift the focus from treating disease to promoting long-term well-being through better food.

*4.3 Future Directions*

Progress in this area will require more complete and harmonized data on both food composition and human biology. Global initiatives such as the Periodic Table of Food Initiative (PTFI) are building molecular-level food databases, but these need to be paired with clinical and metabolic data collected under standardized protocols. Collaboration among nutrition scientists, food scientists, data scientists, and clinicians will be essential to integrate datasets that span agriculture, processing, and health outcomes. New metadata systems should capture not only what is in a food but also where, when, and how it was produced, ensuring traceability from origin to effect.

Future research should also focus on creating interpretable AI frameworks that clearly show how molecular features of food contribute to health impacts. These models should help researchers and policymakers understand which variables matter most, such as dietary composition, processing, or lifestyle and guide targeted interventions. On a societal level, AI-guided nutrition has the potential to reduce chronic disease rates, lower healthcare costs, and promote equitable access to healthy foods. When combined with improved education, labeling, and public awareness, AI can help move the global food system toward one that nourishes both people and the planet.

**5. Education, Awareness, and Training**

*5.1 Background*

Throughout history, technological revolutions have redefined how people learn, work, and produce. The First Industrial Revolution replaced manual labor with mechanized production, creating a demand for basic reading skills and structured factory settings. The Second introduced electricity and assembly lines, expanding the need for skilled workforce and organized management. The Third, or digital revolution, brought computing and automation, pushing education toward programming, data analysis and systems thinking. Today, the Fourth Industrial Revolution, driven by AI, biotechnology, and advanced automation, is transforming every sector at unprecedented connection, speed and scale (Schwab, 2016).

However, the pace of AI integration far exceeds the adaptation of current education and workforce systems. Few academic programs in fields such as agriculture, food science, computer science, and engineering effectively teach students cross-disciplinary data literacy, AI ethics or algorithmic decision-making, leaving them unprepared for interdisciplinary innovation. Meanwhile, experienced industry professionals face similar gaps: those proficient in data analysis lack practical understanding of the food system, while technologists and engineers may struggle to adopt new tools and interpret AI outputs.

Failing to align technological progress with workforce development could widen social and economic disparities, limiting the broader benefits of AI. Closing these gaps will require coordinated action across sectors, which include, developing interdisciplinary curricula, shared digital infrastructure, and ethical governance frameworks to link technical innovation with equitable, human-centered education and training.

*5.2 AI for Education and Workforce Development*

AI is changing how professionals in the food value chain learn, collaborate, and solve problems. In academic settings, AI-enabled learning platforms can simulate real-world food systems, allowing students to explore supply chain, formulation, or processing scenarios in a virtual environment. This promotes systems thinking and builds confidence in applying computational tools to practical challenges. AI can also individualize learning by adjusting instruction pace and content to match a student's background, reinforcing both technical and analytical skills.

As these learners transition into the workforce, the same technologies are reshaping how skills are applied and maintained. In professional settings, AI supports continuous learning and operational training. Intelligent tutoring systems can guide new employees through equipment setup or troubleshooting, linking digital instructions to hands-on activities. Virtual simulations and augmented reality can replicate real manufacturing environments for safe, repeatable training. These tools preserve institutional knowledge by embedding expert experience directly into learning systems. By combining virtual and physical problem-solving, the next generation of workers can develop both cognitive and practical skills essential for a resilient food system.

These shifts in education and professional training highlight the need for a more integrated approach to developing AI literacy. Education, must evolve beyond handling AI as a technical discipline. Food scientists do not need to become computer scientists, but they must understand how to use AI tools responsibly and interpret their outcomes. Likewise, computer science students must learn the fundamentals of food systems to design relevant and interpretable applications. Interdisciplinary courses, co-taught modules, and intensive training formats, such as the AIBridge Boot Camp, where food science students gain skills in programming, model interpretation, and AI ethics, illustrate how short, targeted experiences can accelerate digital literacy and confidence (AIFS, 2025).

*5.3 Future Directions*

The Fourth Industrial Revolution applied to the food industry is not defined solely by technological progress but also by how society adapts and guides it. Its defining features are the fusion of digital, physical, and biological systems, and the speed at which change occurs. In this context, education and workforce development must evolve from static knowledge transfer to continuous adaptation and collaboration. The food and agriculture sectors, long dependent on practical expertise, now require a workforce equally skilled in data and computation systems.

Future progress will depend on aligning education, research, and policy to build an adaptive, interdisciplinary workforce. Curricula should integrate data science and AI fundamentals into food and agricultural programs while maintaining strong grounding in core scientific and engineering principles. Communication across domains, learning to "speak both languages" of computation and application, should be treated as a core competency rather than an additional skill. Industry partnerships will be critical to define evolving skill needs, share training infrastructure, and provide authentic learning environments that mirror real production and processing systems.

AI can also help evaluate and improve education itself. Adaptive assessment tools can test not just knowledge retention but also problem-solving ability, creativity, and reasoning. These systems will help educators measure whether students truly understand what they have learned rather than only memorize information. In the longer term, AI-enabled education should foster curiosity, problem solving, ethical awareness, and a sense of responsibility for the human and environmental impacts of technology. The integration of intelligent digital tools with experiential and value-driven learning will ensure the food sector cultivates a generation of professionals who are not only technologically fluent but also capable of critical thinking, collaboration, and innovation

**Recommendations and Next Steps**

The discussions on the five thematic areas highlighted a common need for integration of data, disciplines, and people. Realizing the potential of AI for food processing will require coordinated action from academia, industry, and government. The recommendations can be summarized in the following:

**Data infrastructure is key.** Developing interoperable standards for food composition, process monitoring, and traceability will enable collaboration while protecting privacy and intellectual property. Shared, high-quality datasets and open benchmarks are essential to validate AI models and compare performance across systems and commodities.

**Research should emphasize transparency and interpretability.** AI should be built on scientific principles that explain, not obscure, how results are generated. Hybrid models that combine physics, chemistry, and biology with data analytics will build confidence and accelerate adoption.

**Education and training must evolve alongside technology.** Academic programs should integrate AI literacy into food and agricultural sciences, while professional development should expand opportunities for hands-on learning. As AI automates routine tasks, preserving pathways for mentorship and apprenticeship will be critical to sustain expertise and ensure future leadership. Human creativity, critical thinking and problem-solving must remain central to training. Technology should be used to enhance human labor rather than replace it.

**Collaboration must be organized.** Public–private partnerships, international networks, and cross-sector initiatives should align around shared digital infrastructure, open standards, and long-term investment in workforce development. The complexity of the food system demands collective leadership rather than isolated initiatives. Building transparent, interoperable, and ethically guided collaborations accelerate innovation, strengthen public trust, and ensure that food systems deliver measurable benefits for people and the planet.

**Conclusion**

Artificial Intelligence is transforming how food is produced, processed, consumed, and understood, connecting science, technology, and human creativity in ways that were previously impossible. The insights from the AI for Food Product Development Symposium make clear that this transformation is not simply technical, but it is cultural, scientific, and ethical. When used responsibly, AI can help create a food system that is predictive instead of reactive, data-informed instead of intuitive, and collaborative instead of fragmented.

The future of food will depend on how effectively these technologies are integrated into practice. Progress will emerge not from replacing human expertise but from enhancing it, using

AI to reveal hidden patterns, optimize processes, and guide better decisions. Ensuring that innovation advances alongside ethics, transparency, education, and shared purpose will allow AI to serve as a tool for both human and planetary well-being. The task ahead is a collective one: to design intelligent systems that sustain life, promote health, and safeguard the environment for generations to come.

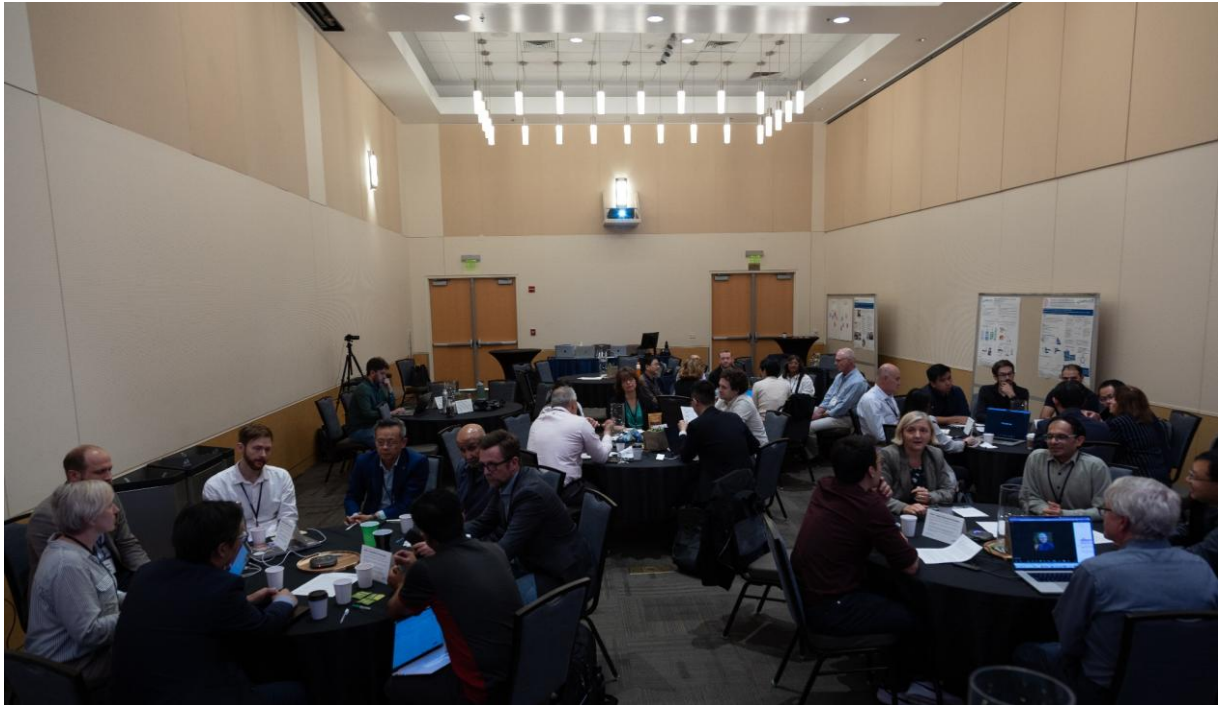

White paper discussion tables at the AI symposium

**AIFPDS white paper discussion participants**

**Adams, Sean**
Professor, UC Davis Health

**Barboza Gardner, Mariana**
Research Program Manager, IIFH, UC Davis

**Beckley, Jacqueline**
Founder and President, The U&I Group

**Bischof, Reto**
Head of Global R&D, Bühler Group

**Bleibaum, Rebecca**
President/Chief of Sensory Intelligence, Dragonfly SCI

**Brown, Steve**
Associate Director of AIFS, Director of AIVO, UC Davis

**Chen, Chang**
Assistant Professor, Cornell University

**Cornyn, Chris**
Multiple Food & Beverage Start-Ups

**Delarue, Julien**
Associate Professor, UC Davis

**Ganjyal, Girish**
Professor, Extension Food Processing Specialist, Washington State University

**Gravelle, Andrew**
Assistant Professor, UC Davis

**Grey, Robert**
Director of Sustainable Future, Plug and Play

**Guinard, Jean-Xavier**
Professor, UC Davis

**Herzog, Leslie**
The U&I Group, UC Davis Food Science and Technology Leadership Board

**Ivanek, Renata**
Professor, Cornell University

**Li, Yonghui**
Associate Professor, Kansas State University

**Lim, Lik Xian**
Ph.D Candidate, Agricultural Chemistry Graduate Group, UC Davis

**Liu, Xin**
Professor, UC Davis

**Mavrich, Travis**
Bioinformatics Technical Lead, PIPA

**Mishra, Ritu**
Associate Research Director, The Clorox Company

**Nguyen, Ha**
Assistant Professor of Sensory Science, UC Davis

**Nitin, Nitin**
Professor, UC Davis

**Prado, Ivor**
Science and Grant Writer, AIFS, UC Davis

**Sablani, Shyam**
Professor, Washington State University

**Sastry, Sudhir**
Professor, Ohio State University

**Schmitz, Harold**
Professor/Partner, UC Davis /TMG

**Shaya, David**
Principal Technologist - Predictive Food Formulation, Ingredion Inc.

**Shrimali, Manmit**
CEO, Turing Labs

**Siegel, Justin**
Professor, Faculty director of IIFH, UC Davis

**Singhasemanon, Kristin**
 Director of Communications and Engagement, AIFS, UC Davis

**Spang, Ned**
Associate Professor, UC Davis

**Stamelos, Chris**
CTO, PIPA

**Tagkopoulos, Ilias**
Professor, Director of AIFS, UC Davis

**Tan, Juzhong**
Assistant Professor, University of Delaware

**Tang, Juming**
Professor & Department Chair, University of Washington

**Tse, Betsy**
Food & Beverage Advisor and Innovator, Turing, Plug & Play

**van Hal, Jolanda**
Writer / Contributor, Nutrition Insights

**Voit, Daniel**
CEO, Blentech

**Watson, Nik**
Professor, University of Leeds, UK

**Weerts, Keith**
VP of Software Development, Blentech

**Xu, Changmou**
Assistant Professor, University of Illinois Urbana-Champaign

**Xue, Jerry**
Chief Engineer, Blentech

**Zhou, Xu**
Postdoctoral Fellow, AIFS, UC Davis